\documentclass[runningheads]{llncs}
\usepackage[T1]{fontenc}
\usepackage{amsmath}
\usepackage{amssymb}
\usepackage{cite}
\usepackage{graphicx}
\usepackage{multirow}
\usepackage{tabularray}
\graphicspath{{images/}}
\begin{document}
\title{Phase Aberration Correction without Reference Data: An Adaptive Mixed Loss Deep Learning Approach}
\titlerunning{PAC without Reference Data: An Adaptive Mixed Loss DL Approach}
\author{Mostafa Sharifzadeh\inst{1,2} \and
	Habib Benali\inst{1,2} \and
	Hassan Rivaz\inst{1,2}}
\authorrunning{M. Sharifzadeh \textit{et al.}}
\institute{Department of Electrical and Computer Engineering, Concordia University\and PERFORM Center, Concordia University, Montreal, QC, Canada.}
\maketitle
\begin{abstract}
Phase aberration is one of the primary sources of image quality degradation in ultrasound, which is induced by spatial variations in sound speed across the heterogeneous medium. This effect disrupts transmitted waves and prevents coherent summation of echo signals, resulting in suboptimal image quality. In real experiments, obtaining non-aberrated ground truths can be extremely challenging, if not infeasible. It hinders the performance of deep learning-based phase aberration correction techniques due to sole reliance on simulated data and the presence of domain shift between simulated and experimental data. Here, for the first time, we propose a deep learning-based method that does not require reference data to compensate for the phase aberration effect. We train a network wherein both input and target output are randomly aberrated radio frequency (RF) data.
Moreover, we demonstrate that a conventional loss function such as mean square error is inadequate for training the network to achieve optimal performance. Instead, we propose an adaptive mixed loss function that employs both B-mode and RF data, resulting in more efficient convergence and enhanced performance.

\end{abstract}
\section{Introduction}
Phase aberration is one of the main sources of image quality degradation in ultrasound imaging caused by spatially varying sound speed while traveling through a heterogeneous medium. This effect alters the focal point in focused imaging and disturbs the flat wavefront propagation in plane-wave imaging during transmission, and prevents coherent summation of echo signals in both imaging modes during the reception.

Many techniques attempted to compensate for the phase aberration effect by utilizing the generalized coherence factor \cite{Li2003}, finding optimal sound speeds \cite{Napolitano2006}, frequency-space prediction filtering (FXPF) \cite{Shin2018a}, determining a local aberration profile at each point \cite{Chau2019}, employing distortion matrix \cite{Lambert2021a}, estimating the spatial distribution of sound speed in a given medium \cite{Sanabria2018a, Raua, Stahli2020, Jakovljevic2018, Brevett2022, Ali2022}, or designing beamformers robust to this artifact \cite{Bendjador2020}.
Recently, various deep learning (DL)-based techniques for phase aberration correction have been proposed, including the estimation of aberration profiles from B-mode images \cite{Sharifzadeh2020}, sound speed distribution from raw RF channel data \cite{Feigin2020a, Jush2020, Young2022} or from in-phase and quadrature (IQ) data \cite{KhunJush2021, Simson2023}, estimation of the aberrated point spread function \cite{Shen2022}, prediction of aberration-free RF data \cite{Koike2023}, and training beamformers robust to this artifact \cite{Luchies2020, Khan2022}.
However, the main challenge in utilizing DL-based approaches is the absence of a precise means of obtaining non-aberrated ground truths and relying solely on simulated data for training, leading to a performance drop when testing on experimental data due to domain shift. Recent studies have recognized the need to eliminate the requirement of ground truths; however, even in such efforts, reconstructed images with a fixed sound speed value of 1540 m/s were still considered clean images \cite{Khan2022}.

In this study, we propose a novel DL-based method for correcting the phase aberration effect in single plane-wave images that does not require ground truth, wherein both input and target output are randomly aberrated beamformed RF data. Moreover, we introduce an adaptive mixed loss function that employs both B-mode and RF data to enhance the performance of the network. We will release both our code and data publicly at *****.

\section{Methodology}
\label{sec:Methodology}

\subsection{Phase Aberration Model and Implementation}
We modeled phase aberrations as near-field phase screens in front of the transducer that introduces delay errors during both transmission and reception. The aberration profile is an array representing delay error values assigned to each transducer element \cite{Sharifzadeh2020}.
We generated random aberration profiles by convolving a Gaussian function with Gaussian random numbers \cite{Dahl2005a}, where their root mean square amplitude, and autocorrelation's full width at half maximum was varied ranging from 20 to 80 ns, and from 4 to 9 mm, respectively. These ranges were chosen to encompass a broad range of tissues according to the reported numbers in the literature \cite{Fernandez2001}.
\\\\
\noindent\textbf{Simulated Aberration.}
\label{sec:Methodology:PhaseAberrationImplementation:SimulatedAberration}
To apply aberration to simulated data, we utilized full synthetic aperture data and synthesized aberrated plane-wave images under linear and steady conditions. Consider a linear array transducer consisting of $N$ elements and oriented such that the $x$-axis is parallel to its length and the $z$-axis represents image depth. After a single plane-waive transmission, the received echo signal at time $t$ by element $n$ located at $x_n$ can be calculated as follows:
\begin{equation}
	\label{eq-RF}
	RF\left(x_n,t\right)=\sum_{m=1}^{N}{RF_{CH}\left(x_m,x_n,t+\tau_a\left(x_m\right)\right)}
\end{equation}
where $RF_{CH}\left(x_m,x_n,t\right)$ is the received echo signal at time $t$ by element $n$ located at $x_n$ solely due to excitation of element $m$ located at $x_m$ and $\tau_a(x_m)$ is the delay error that element $m$ experiences according to the aberration profile. To simulate the aberration effect during transmission, we assumed asynchronous excitation times for elements by applying delay errors $\tau_a(x_m)$ imposed by the aberration profile.
In the absence of phase aberration, the required time for the acoustic wave to travel to point $(x, z)$ and return to the element $n$ located at $x_n$ is:
\begin{equation}
	\tau(x_n,x,z)=(z+\sqrt{z^2+{(x-x_n)}^2})/c 
\end{equation}
where $c$ is the sound speed.
The phase aberration effect in reception was implemented as a set of disordered delay times corresponding to backscattered signals. To this end, and given the calculated time delay, each point $(x,z)$ within the image can be reconstructed as
\begin{equation}
	\label{eq-s}
	s(x,z)=\sum_{n=k-a/2}^{k+a/2}{RF(x_n,\tau(x_n,x,z)+\tau_a(x_n))\ }
\end{equation}
where $k$ is the nearest transducer element to $x$, and aperture $a$ can be expressed using the $f$-number$\ =z/a$. All images in this study were reconstructed using 384 columns along the lateral axis and a $f$-number of 1.75.
\\\\
\noindent\textbf{Quasi-Physical Aberration.}
\label{sec:Methodology:PhaseAberrationImplementation:QuasiPhysicalAberration}
To induce a quasi-physical aberration in an experimental phantom, we programmed a Verasonics Vantage 256 scanner to excite transducer elements asynchronously based on a given aberration profile. Delay errors were written to the \textit{TX.Delay} array to generate an aberrated wavefront during transmission and were also applied during image reconstruction.
\\\\
\noindent\textbf{Physical Aberration.}
An uneven layer of chicken bologna with heterogeneous texture was placed between the probe and the phantom. The thickness of the left and right halves were approximately 3 mm and 6 mm, respectively, and the conductive gel was used to fill the gap between the thinner half and the probe.

\subsection{Datasets}
\label{ssec:Datasets}
\noindent\textbf{Simulated.}
\label{ssec:Datasets:Simulated}
We simulated a dataset comprising 160,000 images with fully developed speckle patterns using Field II \cite{Jensen1996}. Phantoms were 45 mm $\times$ 40 mm and positioned at an axial depth of 10 mm from the transducer, and transducer settings were set based on those of the 128-element linear array transducer L11-5v (Verasonics, Kirkland, WA). The central and sampling frequency were set to 5.208 MHz and 104.16 MHz (downsampled to 20.832 MHz after simulation), respectively.
Initially, 600 images were simulated with anechoic, hypoechoic, and hyperechoic regions (200 samples per type), using segmentation masks taken from XPIE dataset \cite{Xia2017}, as explained in \cite{Sharifzadeh2021}. Then, similar to \cite{Hyun2019a}, 1000 more images were simulated by weighting the scatterers' amplitude according to their respective positions in natural images taken from the XPIE dataset.
Finally, the full synthetic aperture simulations were employed to synthesize 100 randomly aberrated versions of each image according to the procedure elaborated in subsection \ref{sec:Methodology:PhaseAberrationImplementation:SimulatedAberration}.
In addition, a test image comprising two anechoic cysts with diameters of 10 mm and 15 mm, located at central lateral positions and depths of 10 mm and 28 mm, respectively, was simulated using the same configuration. In this case, a non-aberrated version was also synthesized for evaluation purposes. Note that the non-aberrated versions depicted in all figures were solely intended for illustration purposes and were not utilized for training or testing.
\\\\
\noindent\textbf{Experimental Phantom.}
\label{ssec:Datasets:ExperimentalPhantom}
An L11-5v linear array transducer was operated using a Verasonics Vantage 256 system (Verasonics, Kirkland, WA) to scan a multi-purpose multi-tissue ultrasound phantom (Model 040GSE, CIRS, Norfolk, VA). We acquired one scan of anechoic cylinders for testing and an additional 30 scans from other regions of the phantom for fine-tuning. In each acquisition, 51 single plane-wave images were captured at a high frame rate, including one non-aberrated image (only for visualization purposes) and 50 randomly aberrated images as elaborated in subsection \ref{sec:Methodology:PhaseAberrationImplementation:QuasiPhysicalAberration}.

\subsection{Training}
\label{ssec:Training}
Inspired by Lehtinen \textit{et al.}~\cite{Lehtinen2018}, we employed a U-Net \cite{Ronneberger2015} to map beamformed RF data inputs to beamformed RF data target outputs, where both were distinct randomly aberrated versions of the same realization. The network was trained on 1600 simulated samples for 5000 epochs, wherein each sample comprised 100 aberrated versions, and in each epoch, a random pair of aberrated versions were mapped to each other. Images were downsampled laterally by a factor of 2, normalized by dividing them by their maximum value, followed by applying the Yeo-Johnson power transformation. The sigmoid function was employed as the activation function of the last layer, and the batch size was 32. We utilized Adam with a zero weight decay as the optimizer. The learning rate was initially set to $10^{-3}$ and halved at epochs 500, 1000, 1500, and 4000.
Fine-tuning on experimental images was performed with the same configurations by extending the training by an additional 20\% of the original epochs while utilizing a constant and substantially lower learning rate of $5\times10^{-5}$. To deal with the attenuation in experimental images, we partitioned images into three axial sections, each with a 3\% overlap, and fine-tuned a distinct network for each depth. During testing, we fed each depth to its corresponding network, patched the outputs, and blended the envelope of overlapping margins using weighted averaging before displaying the final image. We implemented the method using the PyTorch package and trained and fine-tuned all the models on two NVIDIA A100 GPUs in parallel.
\subsection{Loss Function}
\label{ssec:LossFunction}
Let $\boldsymbol{S}, \boldsymbol{S'}, \boldsymbol{\hat{S}}$ $\in \mathbb{R}^d$ denote input aberrated RF data, target output aberrated RF data, and output of the network, respectively. The aberration-to-aberration problem can be formulated as
\begin{equation}
	\boldsymbol{\hat{S}} = f_{cnn}(\boldsymbol{S}, \boldsymbol{\theta})
	\quad\quad\quad
	\boldsymbol{\theta^{\ast}} = \underset{\theta}{argmin} \; L(\boldsymbol{S'}, \boldsymbol{\hat{S}})
	\vspace{-5pt}
\end{equation}
\noindent where $f_{cnn}: \mathbb{R}^d \rightarrow \mathbb{R}^d$ is the U-Net, and $\boldsymbol{\theta}$ are the network's parameters. During the training phase, an optimizer is utilized to find optimal parameters $\boldsymbol{\theta^{\ast}}$ that minimize the error, measured by a loss function $L$, between network's output $\boldsymbol{\hat{S}}$ and target output $\boldsymbol{S'}$. In this problem, input and target output were highly fluctuating aberrated RF data that were randomly substituted at each epoch. We showed in a pilot study that the network encounters challenges when comparing RF data directly using a conventional mean square error (MSE) loss defined as:
\begin{equation}
	L_{mse}(\boldsymbol{S'}, \boldsymbol{\hat{S}}) = \frac{1}{d}||\boldsymbol{S'}-\boldsymbol{\hat{S}}||^2
\end{equation}
We demonstrated that comparing B-mode data in the loss function resulted in improved network convergence.
\begin{equation}
	L_{b\text{-}mode}(\boldsymbol{S'}, \boldsymbol{\hat{S}}) = \frac{1}{d}||B\text{-}mode(\boldsymbol{S'})-{B\text{-}mode}(\boldsymbol{\hat{S}})||^2
\end{equation}
where $B\text{-}mode(.)$ is the log-compressed envelope data standardized by mean subtraction and division by its standard deviation.

To further improve the network performance, we proposed an adaptively mixed loss function that gradually shifts from B-mode loss to RF loss as the training progresses toward convergence:
\begin{equation}
	L_{adaptive\_mixed}(\boldsymbol{S'}, \boldsymbol{\hat{S}}) = (1-\alpha)L_{b\text{-}mode}(\boldsymbol{S'}, \boldsymbol{\hat{S}}) + \alpha L_{mse}(\boldsymbol{S'}, \boldsymbol{\hat{S}})
\end{equation}
where $\alpha= (current\ epoch\ number)/(total\ number\ of\ epochs)$.

\section{Results and Discussion}
\label{sec:Results}
\begin{figure}[tb]
	\includegraphics[width=\textwidth]{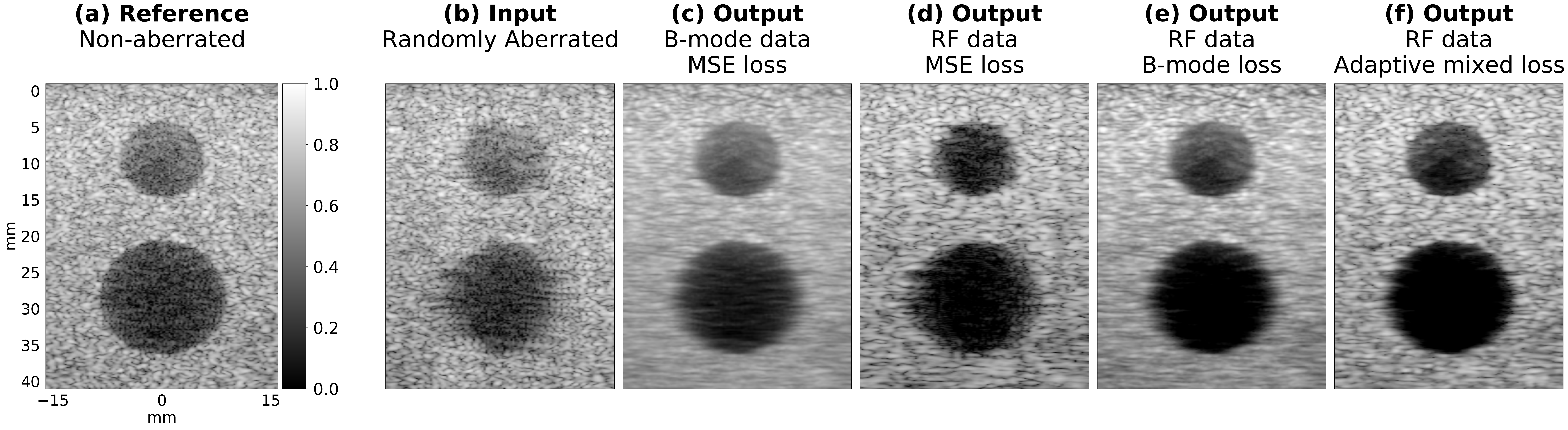}
	\caption{The effect of training using different data types and loss functions. All images are normalized by their maximum intensity and shown on a 50 dB dynamic range.}
	\label{fig-compare-losses}
\end{figure}

\begin{figure}[tb]
	\includegraphics[width=\textwidth]{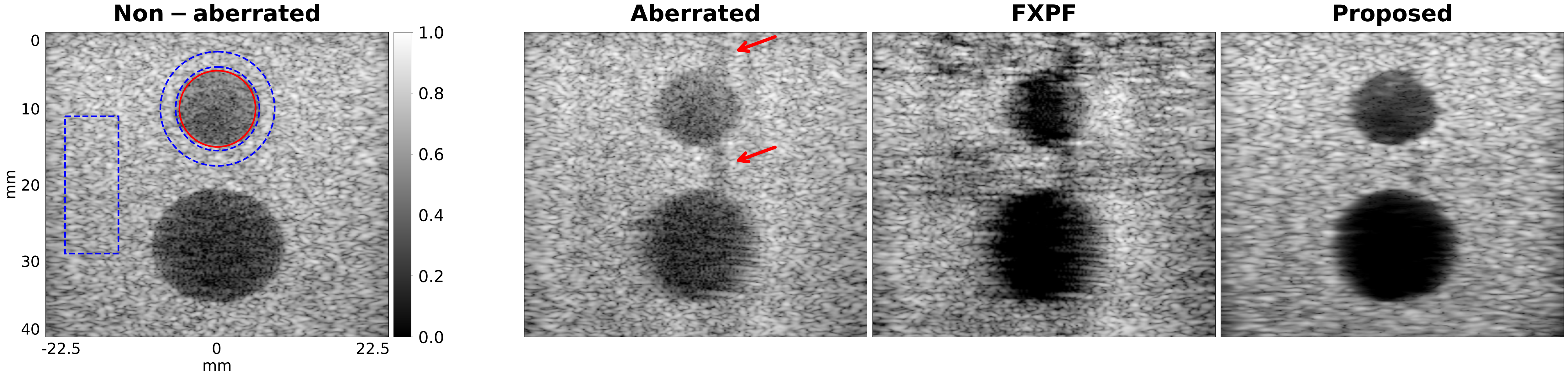}
	\caption{Comparison of methods for the simulated test image. All images are normalized by their maximum intensity and shown on a 50 dB dynamic range.}
	\label{fig-results-simulation}
\end{figure}

\textbf{Pilot Study. }
To assess the performance of the proposed loss function, we conducted an isolated pilot study merely using the simulated test image. The test image consisted of 100 aberrated versions of the same realization, where 99 versions served as a training set, and the remaining one version was used for testing in this pilot study. The network was trained using the configuration specified in Section \ref{ssec:Training}, in which during each epoch, each of the 99 versions was randomly mapped to another version. We trained four different networks, fed them with the test version, shown in Fig. \ref{fig-compare-losses}(b), and evaluated their corresponding output. The first network was trained exceptionally using B-mode data as input and output instead of RF data, with MSE loss. The output is shown in (c), where it is evident that while the cyst borders were mostly recovered, the output appears to be blurry compared to the non-aberrated (a) and aberrated (b) images, which is consistent with the findings reported in \cite{Gobl2022}. Nonetheless, the speckle pattern in the image contains valuable information that can be utilized in applications such as elastography \cite{Tehrani2022}.
 Motivated by the richer information content of RF data, we retrained the network using RF data, resulting in a sharper output, shown in (d). However, the network encountered challenges in recovering cysts borders compared to the B-mode data case.
Inspired by the results from the training with B-mode and RF data, we combined both approaches by retraining the network with RF data but using MSE of B-mode data as the loss function. As shown in (e), this approach exploited the advantages of the rich information in RF data and produced a sharper image compared to (c) while retaining the ability to recover boundaries more efficiently compared to (d).
To further improve the performance, we employed the adaptively mixed loss function. As shown in (f), this approach maintained the ability to recover boundaries while producing a sharper image compared to (e). Our interpretation is that this loss function guides the optimizer towards a correct solution by initially utilizing simpler data, before gradually incorporating more complex, fluctuating RF data to take full advantage of the richer information, like curriculum learning \cite{Bengio2009}. This helps to avoid getting stuck in local minima during the initial stages of the optimization.
\\\\
\noindent
\textbf{Main Study. }
Here, we utilized the adaptive mixed loss function and trained a network using 1600 simulated images, and evaluated its performance on 100 aberrated versions of the test image. One such aberrated version is shown in Fig. \ref{fig-results-simulation}. We compared the proposed method with the FXPF method with an autoregressive model of order 2, stability factor of 0.01, kernel size of one wavelength, and 3 iterations. These settings were chosen based on our observation that they yielded the best results in our case. Two red arrows in the aberrated image highlight the shadowing effect resulting from the perturbed wavefront due to aberration during transmission. The proposed method demonstrated superior performance compared to FXPF, which failed to detect and correct this effect due to its reliance solely on local signal information of a single image. Note that in contrast to the pilot study, in this case, the network had never seen a perfect circular cyst during the training phase.

To perform a quantitative evaluation, we calculated contrast, generalized contrast-to-noise ratio (gCNR), and speckle signal-to-noise ratio (SNR) metrics for both cysts in the test image. The results were obtained by averaging across 100 aberrated versions and reported in Table \ref{tbl1}(a). Target and background regions used for calculating metrics are demonstrated for the top cyst in the non-aberrated image in Fig. \ref{fig-results-simulation}, where the target refers to inside a circle with the same radius and center as the cyst (red circle). For contrast and gCNR, the background refers to a region between two concentric circles with radii of 1.1 and 1.5 times the cyst radius (blue circles), while for speckle SNR, it refers to a rectangle region far from the cysts (blue rectangle). For the bottom cyst, the regions were selected in a similar way. Metrics were calculated on the envelope detected image in the linear domain and prior to applying log-compression.

\begin{figure}[tb]
	\centering
	\includegraphics[width=\textwidth]{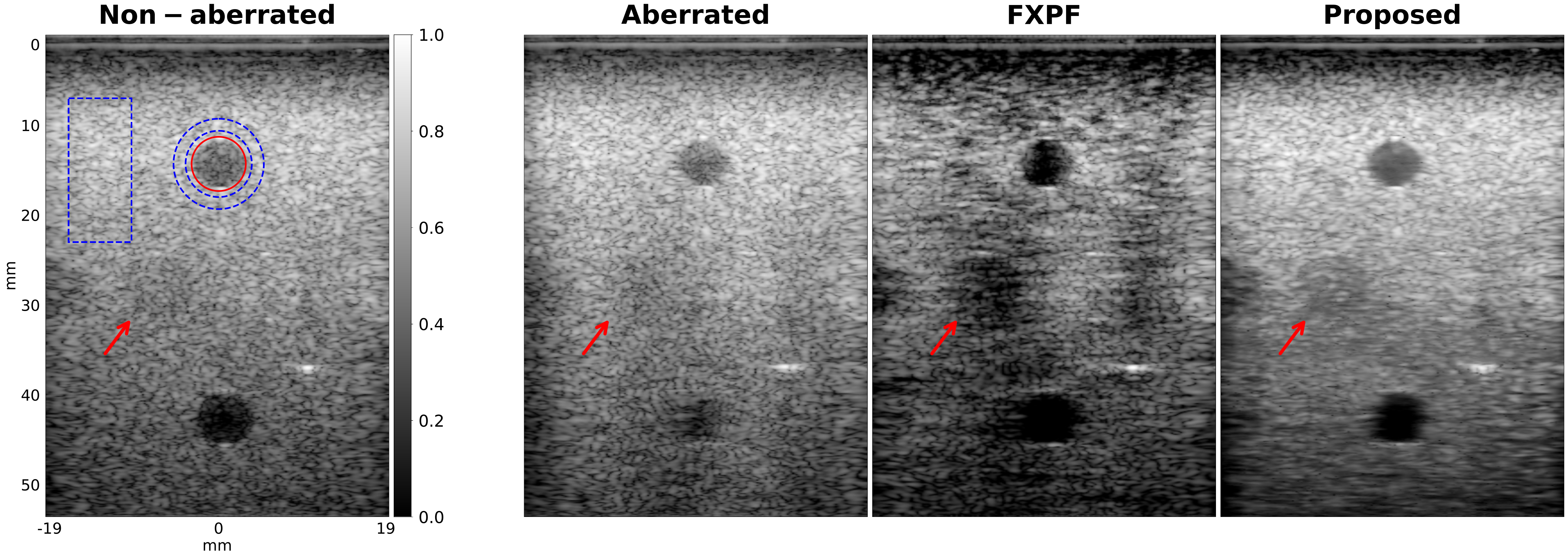}
	\caption{Comparison of methods for the experimental phantom with quasi-physical aberration. All images are normalized by their maximum intensity and shown on a 60 dB dynamic range.}
	\label{fig-results-numerical_phantom}
\end{figure}

Fig. \ref{fig-results-numerical_phantom} presents the results for one of the aberrated versions of the experimental phantom test image, which was acquired with quasi-physical aberrations. It is evident that the proposed method compensated for the aberration effect without introducing unwanted artifacts. Notably, the proposed method outperformed FXPF by recovering the borders of the bottom cyst more accurately. Moreover, the cyst indicated with the red arrow was recovered with a more precise contrast, aligned with our prior knowledge that it was a -3 dB hypoechoic cyst. The quantitative metrics obtained for the top and bottom anechoic cysts were calculated, averaged over all 50 aberrated versions, and presented in Table \ref{tbl1}(b).

\begin{figure}[tb]
	\centering
	\includegraphics[width=0.87\textwidth]{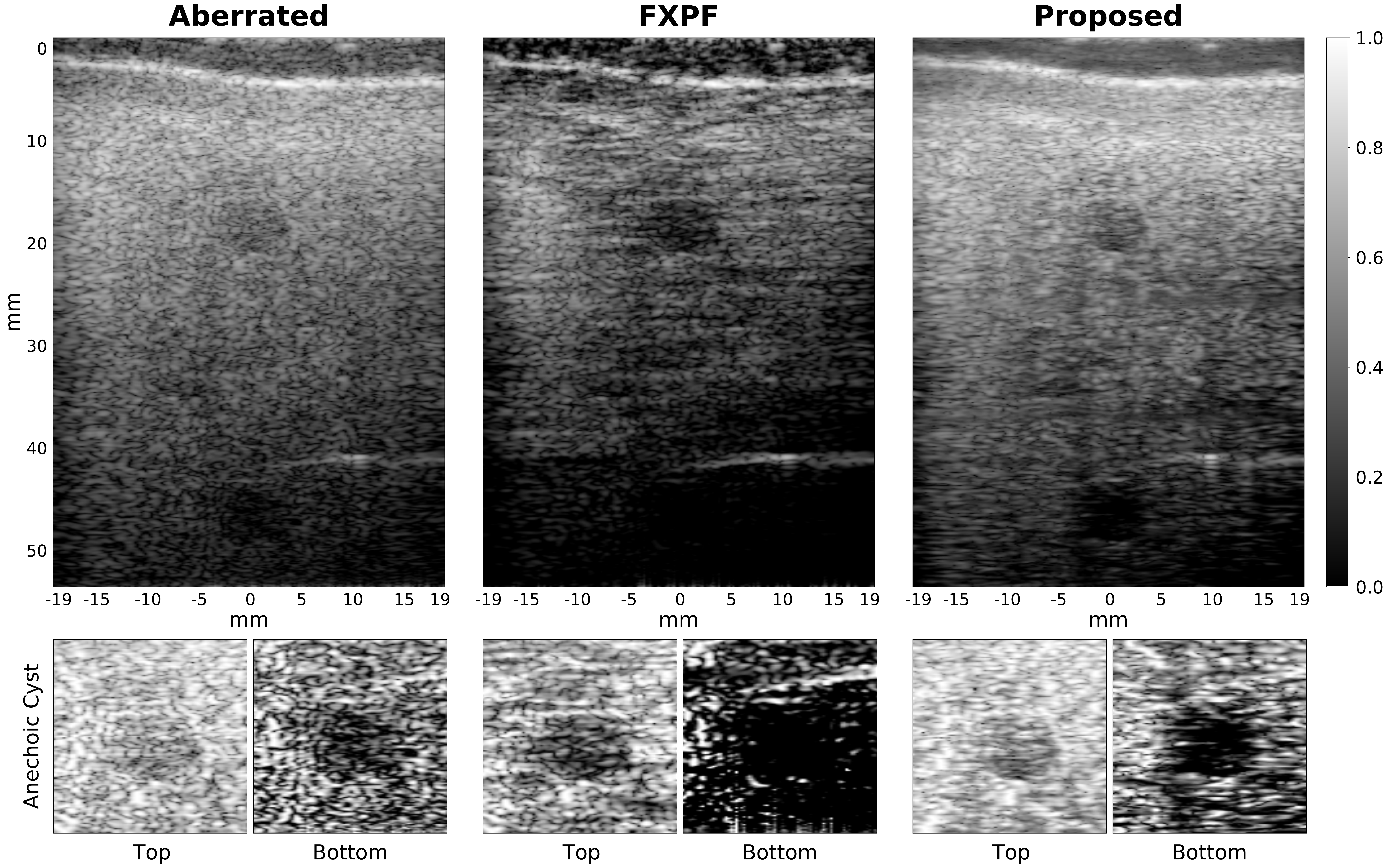}
	\caption{Comparison of methods for the experimental phantom aberrated with a physical aberrator layer. Images in the first row are shown on a 60 dB dynamic range.}
	\label{fig-results-physical_phantom}
\end{figure}

\noindent Fig. \ref{fig-results-physical_phantom} presents the results for the experimental phantom aberrated with a physical aberrator. To enhance visual comparability, the top and bottom cysts were cropped and displayed under their corresponding images, where each horizontal line of the cropped images was normalized to its maximum intensity. While the FXPF method provided a higher contrast for the top cyst, it resulted in an overestimation of its size. In contrast, the proposed method recovered the true cyst borders more accurately, particularly for the bottom cyst.

\begin{table}[tb]
	\centering
	\caption{Quantitative measures for (a) the simulated test image and (b) the experimental phantom with quasi-physical aberration.}
	\label{tbl1}
	\begin{tblr}{
			cell{1}{2} = {l},
			cell{1}{3} = {c},
			cell{1}{4} = {c},
			cell{1}{5} = {c},
			cell{2}{1} = {r=3}{},
			cell{5}{1} = {r=3}{},
			hline{1-2,5,8} = {2-5}{},
		}
		& \textbf{Metric} & Aberrated        & FXPF             & \textbf{Proposed} \\
		\textbf{(a)} & Contrast (dB)   & 14.77 $\pm\ 0.96$ & 21.91 $\pm\ 1.23$ & 24.72 $\pm\ 0.61$ \\
		& gCNR            & 0.84 $\pm\ 0.03$  & 0.85 $\pm\ 0.03$  & 0.96 $\pm\ 0.01$  \\
		& Speckle SNR     & 1.78 $\pm\ 0.08$  & 1.57 $\pm\ 0.10$  & 1.79 $\pm\ 0.08$  \\
		\textbf{(b)}  & Contrast (dB)   & 9.48 $\pm\ 1.24$  & 11.59 $\pm\ 1.44$ & 12.22 $\pm\ 0.44$ \\
		& gCNR            & 0.73 $\pm\ 0.05$  & 0.69 $\pm\ 0.08$  & 0.87 $\pm\ 0.01$  \\
		& Speckle SNR     & 1.61 $\pm\ 0.03$  & 1.13 $\pm\ 0.11$  & 1.62 $\pm\ 0.04$  
	\end{tblr}
\end{table}
\section{Conclusion}
\label{sec:Conclusion}
We proposed a new approach for correcting phase aberration that eliminates the requirement for ground truths. The proposed method employs an adaptive mixed loss function to train a network capable of mapping aberrated RF data to aberrated RF data. This approach permits the utilization of experimental images for training or fine-tuning, without any prior assumptions about the presence or absence of aberration. Furthermore, our experiments demonstrated the feasibility of acquiring data for this method using a programmable transducer, which can acquire multiple aberrated versions of the same scene during a single scan.
 \bibliographystyle{splncs04}

\end{document}